\documentclass{pasj00}
\draft
				
\newcommand{\eg}{e.g., }

\newcommand{\Msun}{M_{\odot}}
\newcommand{\kms}{km~s$^{-1}$}

\def\gsim{\mathrel{\rlap{\lower 4pt \hbox{\hskip 1pt $\sim$}}\raise 1pt
\hbox {$>$}}}
\def\lsim{\mathrel{\rlap{\lower 4pt \hbox{\hskip 1pt $\sim$}}\raise 1pt
\hbox {$<$}}}

\begin{document}
\SetRunningHead{Yamanaka et al.}{Early Spectroscopy of SN 2006X}
\Received{2008 August 17}
\Accepted{2009 April 13}

\title{Early Spectral Evolution of the Rapidly Expanding Type Ia SN 2006X}

\author{Masayuki \textsc{Yamanaka}\altaffilmark{1,2,3},
        Hiroyuki \textsc{Naito}\altaffilmark{4},
        Kenzo  \textsc{Kinugasa}\altaffilmark{5},
        Naohiro \textsc{Takanashi}\altaffilmark{6}, \\
        Masaomi \textsc{Tanaka}\altaffilmark{7},
        Koji S. \textsc{Kawabata}\altaffilmark{2},
        Shinobu  \textsc{Ozaki}\altaffilmark{8},
        Shin-ya \textsc{Narusawa}\altaffilmark{4}, \\
        and Kozo \textsc{Sadakane}\altaffilmark{3},}

\altaffiltext{1}{Department of Physical Science, Hiroshima University, Kagamiyama 1-3-1, Higashi-Hiroshima 739-8526}
\altaffiltext{2}{Hiroshima Astrophysical Science Center, Hiroshima
University, Higashi-Hiroshima, Hiroshima 739-8526}
\altaffiltext{3}{Astronomical Institute, Osaka Kyoiku University, Asahigaoka,  Kashiwara-shi, Osaka 582-8582}     
\altaffiltext{4}{Nishi-Harima Astronomical Observatory, Sayo-cho, Hyogo, 679-5313}
\altaffiltext{5}{Gunma Astronomical Observatory, Takayama, Gunma 377-0702}
\altaffiltext{6}{National Astronomical Observatory of Japan, Mitaka 181-8588}
\altaffiltext{7}{Department of Astronomy, School of Science, University
 of Tokyo, Bunkyo-ku, Tokyo 113-0033}
\altaffiltext{8}{Okayama Astrophysical Observatory, National Astronomical Observatory of Japan, Kamogata, Asakuchi-shi, Okayama 719-0232}
\email{myamanaka@hiroshima-u.ac.jp}


\maketitle

\begin{abstract}
 We present optical spectroscopic and photometric observations of Type Ia
 supernova (SN) 2006X from --10 to +91 days after the $B$-band maximum.
 This SN exhibits one of the highest expansion velocity ever published 
 for SNe Ia. At premaximum phases, the spectra show strong and broad features 
 of intermediate-mass elements such as Si, S, Ca, and Mg, 
 while the O~{\sc i}~$\lambda$7773 line is weak.
 The extremely high velocities of Si~{\sc ii} and S~{\sc ii} lines and
 the weak O~{\sc i} line suggest that an intense 
 nucleosynthesis might take place in the outer layers, favoring a 
 delayed detonation model.
 Interestingly, Si~{\sc ii}~$\lambda$5972 feature is quite shallow, 
 resulting in an unusually low depth ratio of Si~{\sc ii}~$\lambda$5972 
 to $\lambda$6355, $\cal R$(Si~{\sc ii}).
 The low $\cal R$(Si~{\sc ii}) is usually interpreted as a high photospheric  
 temperature.
 However, the weak Si~{\sc iii}~$\lambda$4560 line suggests a low temperature,
 in contradiction to the low $\cal R$(Si~{\sc ii}).
 This could imply that the Si~{\sc ii}~$\lambda$5972 line might be 
 contaminated by underlying emission.
 We propose that $\cal R$(Si~{\sc ii}) may not be a good 
 temperature indicator for rapidly expanding SNe Ia 
 at premaximum phases.
\end{abstract}

\section{Introduction}
Type Ia supernovae (SNe Ia) are thermonuclear explosions of C+O white dwarfs. It is thought that the explosion is triggered 
when the mass of the white dwarf approaches 
the Chandrasekhar limit ($\sim$1.4$\Msun$). 
Their peak luminosity, calibrated using the correlation with the light curves and/or color curves,
is quite uniform, and thus, SNe Ia have been used as ``standard candles'' for measuring the extragalactic distances 
\citep{Phillips1993,Phillips1999,XWang2005}.
Observations of distant SNe Ia led 
to the discovery of the accelerating
expansion of the universe (\cite{Riess1998}; \cite{Perlmutter1999}). 

Spectroscopic properties and their time evolution 
had been thought to be homogeneous in SNe Ia 
(e.g., \cite{Branch1993,Filippenko1997}).
However, recent spectroscopic observations show that the spectra of 
SNe Ia at premaximum phases show diversity in their line velocities and 
profiles \citep{Fisher1997, Benetti2004, Mazzali2005a,
Mazzali2005b, Quimby2006b, Altavilla2007, Garavini2007, Phillips2007,
Sahu2008, XWang2009}.
In particular, the Doppler velocity of Si~{\sc ii} $\lambda$6355 
feature exhibits a large dispersion, and 
the velocity is not correlated with the peak luminosity (\cite{Hatano2000}). 

\citet{Benetti2005} collected some well-observed SNe Ia sample 
and divided them into three groups according to the luminosity
and the temporal gradient of the Si~{\sc ii} velocity.
SNe Ia with both high velocity gradient (HVG) and low velocity
gradient (LVG) have normal luminosity, while the FAINT group is 
dimmer than the two classes above, as 
represented by subluminous SN Ia 1991bg \citep{Filippenko1992}.
\citet{Branch2006} studied near-maximum spectra of SNe Ia and defined 
four groups according to the equivalent width ratio of 
Si~{\sc ii}$\lambda$6355 and Si~{\sc ii}$\lambda$5972 and also
the profile of the Si~{\sc ii}~$\lambda$6355 line near the maximum light.
These studies show that the physics of SNe Ia 
cannot be represented by a family of one parameter such as the luminosity
decline rate $\Delta m_{15}(B)$
\footnote{The difference between the $B$-band magnitudes
at maximum and that at 15 days after the maximum \citep{Phillips1993}.} \citep{Benetti2004}.

\citet{Nugent1995} found that the depth ratio of Si~{\sc ii}~$\lambda$5972 
to Si~{\sc ii}~$\lambda$6355, $\cal R$(Si~{\sc ii}), is correlated with
the absolute luminosity of SNe Ia at maximum brightness.
It is thought that $\cal R$(Si~{\sc ii}) is an 
indicator of the temperature of the photosphere.
At the premaximum phases, $\cal R$(Si~{\sc ii}) 
shows considerable diversity.
\citet{Benetti2005} showed that, at the premaximum phases,
$\cal R$(Si~{\sc ii}) in HVG SNe is high in the earliest phases
and declines rapidly, while that of LVG SNe does not change 
significantly.
\citet{Tanaka2008} suggested that the photospheric temperature 
in HVG SNe is lower than that of LVG SNe at premaximum phases
($\gsim 1$ week before maximum) from their spectrum analysis. \\\\\\

SN~2006X ($\alpha_{2000}=12^{h} 22^{m}
53^{s}.90$, $\delta_{2000}=-15^\circ 48' 32.9''$) was
discovered on 2006 February 4.75 UT by S. Suzuki and 
M. Migliardi (2006), independently, near the center 
of the nearby galaxy NGC~4321 (M100).
The distance to NGC~4321 is derived as $\mu =30.91\pm 0.14$ 
by the Cepheid calibration of the Hubble Space Telescope Key Project 
\citep{Freedman2001}.
\citet{Quimby2006a} classified SN~2006X as Type Ia.
They reported that the Si~{\sc ii}~$\lambda$6355 line velocity of this SN
was very high ($20,700$ \kms ) on February 8.35 
($t=-11.6$ d, hereafter, $t$ denotes days from the $B$-band maximum, MJD$=$53785.67. See \S 2.1.).
\citet{XWang2008a} presented extensive optical and NIR 
photometric observations. They derived maximum absolute magnitude 
$M_{V,{\rm max}}=-19.06\pm 0.17$, decline rate of the $B$-band
light curve $\Delta m_{15}(B)=1.17\pm 0.05$, 
and maximum bolometric luminosity
$L_{\rm bol,max}\simeq 1.0\times 10^{43}$ erg s$^{-1}$,
suggesting that this SN is a normally luminous SN Ia.
\citet{XWang2008a} obtained spectra around and after the maximum, 
and suggested that SN 2006X is characterized by strong, highly 
blueshifted absorption lines of intermediate-mass elements.
The Si~{\sc ii}~$\lambda$6355 absorption is very deep and broad
at one week after the maximum, similar to that in
SNe 1984A and 2002bo \citep{Branch2008}.
\citet{Patat2007} reported the detection of the
temporal variation of Na~{\sc i}~D 
absorption line and suggested the presence of dense
circumstellar materials (CSM) around SN 2006X. The possible detection of an inner echo might be also consistent with the
circumstellar dust scenario \citep{XWang2008b}.

In this paper, we present spectroscopic and photometric observations of 
SN~2006X. In particular, we show optical spectra at premaximum phases, 
which are not presented by \citet{XWang2008a}.
The premaximum spectra are useful to explore the physical 
properties in the outermost layers of the SN ejecta.
Our observations and data reduction are described in \S 2.
Results are shown in \S 3.
We discuss the spectroscopic properties of SN~2006X 
at the premaximum phases in \S 4, and finally give conclusions in \S 5.

\section{Observations and Data Reduction}

\subsection{Photometric Observations}

We performed $BVR_{c}I_{c}$ photometric observations on 24 nights from
$t=-10$ d to $+91$ d with two telescopes;
a 0.6m reflector equipped with an ST-9 imager at Nishi-Harima Astronomical
Observatory (NHAO) and a 0.51m telescope equipped with a CCD camera at
Osaka Kyoiku University (OKU). 
Data reduction was performed in a standard manner for
aperture photometry using $IRAF$
\footnote{ $IRAF$ is distributed by the National Optical Astronomy
Observatories, operated by the Association of Universities for Research
in Astronomy, Inc., under contract to the National Science Foundation of the United States. }. 
Although there is a star located 
at $1''$ east and $8''$ north of the SN, it is faint ($V$$=$$17.05$, \cite{XWang2008a}), thus our photometry is not significantly
affected by this star except for the measurements at the last
two epochs. We performed the PSF photometry for the last 
two-epoch images, and derived the magnitudes that are
consistent with those in\citet{XWang2008a}, within $2$ $\sigma$
error. The photometric calibration was performed using the 
comparison stars in \citet{XWang2008a}.
A summary of our photometric observations is shown in table 1.

From our data, we derived the date of $B$-band maximum at MJD$=53785.1$ $\pm$ $1.8$
and maximum magnitude $m_{B,max}$ $=$ 15.30 $\pm$ 0.09 by a polynomial fitting.
Both are consistent with those in \citet{XWang2008a},
who derived the $B$-band maximum date and magnitude as
$53785.67\pm 0.35$ and $15.40\pm 0.03$, respectively.
In this paper, we assume MJD$=53785.67$ as $t=0$ d for 
the convenience of comparison with their study.

\subsection{Spectroscopic Observations}

Spectroscopic observations were performed on 22 nights from $t=-10$ d to
$+84$ d.
18 spectra were taken at NHAO, including four at premaximum phases, with the 2m NAYUTA telescope and a low-resolution
spectrograph, MALLS (Medium And Low-dispersion Long-slit Spectrograph; \cite{Ozaki2005}). 
The typical exposure time is 1800 s.
The wavelength coverage is 4000--6800 \AA\ and the wavelength
resolution is $\sim$17 \AA\ , corresponding to a velocity 
resolution of 850 \kms\ at 6000 \AA. 
Four spectra were taken at Gunma Astronomical Observatory (GAO), including two at premaximum phases,
with the 1.5m telescope equipped with a low-resolution spectrograph, 
GLOWS (Gunma LOW resolution Spectrograph and imager).
The wavelength coverage is 4000--8000 \AA\ and the  
wavelength resolution is $\sim$15 \AA\ (750 \kms).
One late-time spectrum (at $t=63$ d) was obtained
by using the 8.2 m Subaru Telescope (National Astronomical Observatory of Japan) and FOCAS (Faint Object Camera And Spectrograph; \cite{Kashikawa2002}), with the wavelength coverage being 4000--9000 \AA\ and the wavelength resolution being $\sim$14 \AA\ (700 \kms). 
The log of our spectroscopic observations is shown in table 2. 

Data reduction was performed by the standard procedure for
long-slit spectroscopy with $IRAF$ tasks.
Sky background was subtracted by 1D interpolation along the 
focal slit direction. 
The wavelength calibration was obtained by using night-sky
lines taken in the object frames (for NHAO data; \cite{Iye1991}) 
or comparison lamp data (for the others). 
The flux was calibrated with the data of spectrophotometric standard 
stars obtained on the same night. 

In determining the central wavelength of absorption lines, 
we used the method described in \citet{Hachinger2006}.
We first estimated the center of each feature by eyes.
Then, we performed 1D Gaussian fitting to the feature several times,
and derived the mean central wavelength and its standard deviation.
The final uncertainty of the center wavelength was taken as the
root sum square of the standard deviation and the wavelength resolution described above.

\section{Results}

\subsection{Light Curves and Color Curves}

We present light curves and color curves in figures 1 and 2,
respectively. 
The light curves are compared with that of SN~1994D 
(thin lines, \cite{Patat1996}).
From our data, the light curve decline parameter 
is derived as $\Delta m_{15}(B)=1.2\pm 0.1$, which is consistent with
$\Delta m_{15}(B)=1.17 \pm 0.05$ derived by \citet{XWang2008a}.
This value is typical for a normal SN Ia (e.g., \cite{Phillips1999}).

We derived color excesses as $E(B-V)=1.26 \pm 0.17$,
$E(V-R)=0.64 \pm 0.13$, and $E(V-I)=1.23 \pm 0.26$ using the
template color indices given by \citet{Nobili2008}. 
These values are also consistent with 
those by \citet{XWang2008a} within 1$\sigma$ error. 
In figure 2, $B-V$, $V-R$ and $V-R$ curves are
compared with those of other SNe Ia, SNe~2002bo
\citep{Krisciunas2004}, 2002er \citep{Pignata2004}, 2003cg
\citep{Elias2006}, and 2003du \citep{Stanishev2007}. 
The color curves of these SNe are shifted arbitrary 
to match those of SN 2006X.
We find no significant difference between them.

\subsection{Spectral Evolution}

The complete spectral evolution of SN~2006X spectra from 
$t=-10$ d to $+84$ d is shown in figure 3. 
The spectra have been corrected for the redshift of the host
galaxy (v $=$ $+$1,571 \kms\ ; (\cite{Rand1995}).
We show spectra at $t=-10$ d and $-6$ d in figures 4
and 5, respectively, compared with those of other SNe Ia 
at similar phases \footnote{ We took the 
spectra from the SUSPECT database; 
\\ http://bruford.nhn.ou.edu/\~{}suspect/index.html. }. 
For the reddening correction, 
we assumed a color excess of $E(B-V) =1.42$ and an extinction 
coefficient of $R_{V} = A_{V}/E(B-V) = 1.48$ \citep{XWang2008a}. 
Line identifications are given by comparison with well-studied
SNe Ia~2003cg
\citep{Elias2006}, 2002er \citep{Kotak2005}, 2002bo
\citep{Benetti2004}, and 2003du \citep{Stanishev2007}.

At $t=-10$ d (figure 4), the absorption of the 
Si~{\sc ii}~$\lambda$6355 line has the minimum at 5950 \AA, 
corresponding to an expansion velocity of about 19,000 \kms. 
This is one of the highest expansion velocity 
that has ever been observed for SNe Ia at similar phases.
The velocity measured from Si~{\sc ii}~$\lambda$5972
is found to be as high as 17,000 \kms.
The W-shaped S~{\sc ii}~$\lambda$5468 and S~{\sc ii}~$\lambda$5640
(the blend of $\lambda$5612 and $\lambda$5654) 
lines also show larger blueshifts than those of other 
SNe Ia shown in figure 4.
The features around 4700 \AA\ and 4200 \AA\
are also broad, and largely blueshifted.
The Si~{\sc iii}~$\lambda$4560 line is not strong, 
which may be common in the fast-expanding SNe Ia \citep{Pignata2008}. 
In the $t=-9.8$ d GAO spectrum (figure 3), 
a weak O~{\sc i}~$\lambda$7773 line can be seen around 7300 \AA.
A broad absorption trough is also seen around 7800 \AA .
If this trough is produced by the Ca~{\sc ii}~IR triplet, 
the expansion velocity reaches $\gtrsim 26,000$ \kms .

At $t=-6$ d (figure 5), the blueshift of the absorption features is still large.
The Si~{\sc ii}~$\lambda$6355 absorption remains deep and broad. 
In contrast, the Si~{\sc ii}~$\lambda$5972 feature becomes
unusually weak.
The blueshift of the W-shaped sulfur feature is 
still larger than those in the comparison SNe Ia. 
We noticed that the Si~{\sc iii}~$\lambda$4560 feature becomes
stronger than it appeared at $t=-10$ d (figure 4, see \S 4.2).

The strong absorption features of 
intermediate-mass elements such as Si, S, Ca, and Mg
maintained throughout the premaximum phases, while
the O~{\sc i} $\lambda$7773 is very weak at $t=-10$ d 
and almost absent at $t=-1$ d. There is no clear evidence of
carbon feature in the spectra of SN 2006X. 
Implications of these observational properties will be
discussed in \S 4.

The spectral evolution around and after maximum 
has been studied by \citet{XWang2008a}.
The line profiles of the spectra between SN 2006X and
other SNe Ia in comparison become similar when entering 
the nebular phase (figure 3). 
It is noted that the light echo component, which has been found by later phase observations ($t\sim 300$ d) by HST and Keck 
\citep{XWang2008b}, seems to be negligible during our 
observational period because there is no excess in the 
light curve
(see figure 2 in \cite{XWang2008b}).


\section{Discussion}

\subsection{Extremely Large Expansion Velocity
and Constraints on Explosion Model}

In figure 6, we present the temporal evolution of the line 
velocities of Si~{\sc ii}$\lambda$6355 and S~{\sc ii} $\lambda$5640, together with those of other four SNe Ia.
It clearly shows that the velocities of SN 2006X are 
the highest among all the SNe in comparison 
from premaximum phases to $t=+30$ d.
Especially, the Si~{\sc ii} $\lambda$6355 line velocity 
at $t=-10$ d reaches 19,000 \kms.
For example, the expansion velocity of
SN 2006X was found to be higher than that of the normal SNe Ia
by $\gsim$ 5000 \kms during the premaximum phase and
higher than that of SN 2002bo by 2000-3000 \kms.
Such a high-velocity behavior were also observed in
SNe~1983G (\cite{Benetti1991};
\cite{McCall1984}), 1984A \citep{Barbon1989}, 2002bf
\citep{Leonard2005}, 2002bo \citep{Benetti2004}, 2002dj
\citep{Pignata2008}, and 2004dt \citep{Altavilla2007}.
SN 2006X may be one of them with the 
highest expansion velocity.
The velocity of the S~{\sc ii} $\lambda$5640 lines have also
been regarded as a better tracer of the photospheric velocity
\citep{Patat1996}.
Inspection of figure 6 similarly reveals a photospheric 
expansion velocity is very high in SN 2006X. 

The premaximum spectra can
provide constraints on the explosion models.
In deflagration models \citep{Nomoto1984,Roepke2007}, 
elements heavier than Mg are not synthesized in the 
outer layers with $v$ $\gsim$ 15,000 \kms.
In the case of SN~2006X, the Si and S line velocities are 
higher than 15,000 \kms, indicating that these intermediate-mass 
elements are located in such outer layers. 
Thus, the deflagration models may fail to explain the 
high velocity lines of intermediate-mass elements 
in the premaximum spectra.

In delayed detonation models \citep{Khokhlov1991},
an intense nucleosynthesis takes place in the outer layers
and the C+O white dwarf is almost completely burned.
For example, in the CS15DD2 model in \citet{Iwamoto1999},
a Si-rich layer with $X({\rm Si})\gsim 0.1$ extends to 20,000 \kms.
Thus the delayed detonation model may account for the high
velocity absorption features of intermediate-mass elements
seen in SN 2006X.
The weak O~{\sc i}~$\lambda$7773 line is consistent with 
such a scenario.


Another possible scenario of the increasing apparent line 
velocity is the interaction of the SN ejecta with the CSM.
In SN~2006X, the presence of the CSM around the progenitor has
been suggested by \citet{Patat2007} and \citet{XWang2008b}.
The CSM may be stripped-off atmosphere from the companion star
by white dwarf wind during the pre-explosion phase \citep{Hachisu2008}.
The interaction between the SN ejecta and the CSM could produce
the high velocity Ca~{\sc ii} lines \citep{Gerardy2004}
and may also affect the Si~{\sc ii} lines \citep{Tanaka2006}.
Thus, the CSM interactions could also explain the extremely high
Si~{\sc ii} and S~{\sc ii} line velocities
\footnote{
If the evolution of the Si~{\sc ii} and Ca~{\sc ii} lines are strongly
correlated, CSM interaction scenario may be favored.
However, due to the absence of the spectra covering Ca~{\sc ii} lines
(Ca~{\sc ii} H\&K or IR triplet lines),
we refrain from further discussion.
}.

\subsection{Line Depth Ratio $\cal R$(Si~{\sc ii})} 

We discuss the time evolution of the depth ratio of 
the two Si lines, $\cal R$(Si~{\sc ii}). 
The Si~{\sc ii} lines of SN~2006X show unique properties.
The Si~{\sc ii}~$\lambda$6355 feature was very strong
before and around maximum, 
while the Si~{\sc ii}~$\lambda$5972 was visible in the 
earliest spectra but became subsequently rather weak
at around the maximum light.
\citet{Nugent1995} defined the depth ratio as
\begin{equation}
 { \cal R }({\rm Si ~{\sc II} } ) = 
 {\displaystyle \frac{D({\rm Si~{\sc II}}\lambda 5972)}
 {D({\rm Si~{\sc II}}\lambda 6355)} } \mbox{,}
\end{equation} 
where ${\it D}$ is the depth of the absorption feature 
below the pseudo-continuum level.
This spectral index is easy to be applied because it 
does not depend on the reddening correction.
The $\cal R$(Si~{\sc ii}) around the maximum is considered 
to be an indicator of luminosity and/or temperature for SNe Ia: 
a SN with lower $\cal R$(Si~{\sc ii}) is more luminous and 
shows a higher photospheric temperature \citep{Nugent1995,Bongard2006}.
This is explained by the temperature-sensitive Si~{\sc ii}~$\lambda$5972 line,
which becomes stronger as the photospheric temperature is
lower \citep{Hachinger2006,Hachinger2008}.
The time evolution of the $\cal R$(Si~{\sc ii}) during 
the premaximum phases shows diversity among SNe Ia \citep{Benetti2005}.
During the premaximum phases, $\cal R$(Si~{\sc ii}) of HVG SNe
decrease while that of LVG SNe stays nearly constant.
Around the maximum, the $\cal R$(Si~{\sc ii}) of HVG and LVG SNe become comparable.

The decline rate of the Si~{\sc ii}~$\lambda$6355 
line velocity in SN~2006X from $t=0$ d to $+28$ d is 
estimated as $\dot{v}\simeq -130$
\kms\ day$^{-1}$, which puts this object into the HVG group, similar to
SNe~2002bo and 2002er (figure 6).
In figure 7, we show the time evolution of $\cal R$(Si~{\sc ii}). 
The rapidly decreasing trend of $\cal R$(Si~{\sc ii}) in SN~2006X from 
$t=-10$ d to $-6$ d is consistent with SN 2002bo.
However, the value of $\cal R$(Si~{\sc ii})$\simeq 0.1$ 
is unusually low compared to SNe 2002bo and 2002er 
(although we cannot reject the possibility that $\cal R$~(Si~{\sc ii}) 
was much higher at earliest epochs, i.e., $t\lsim -10$ d).
The low $\cal R$~(Si~{\sc ii}) is usually interpreted as a high photospheric temperature.

Si~{\sc iii}~$\lambda$4560, as another temperature indicator, is
expected to be very strong in a SN Ia with high photospheric 
temperature \citep{Benetti2004}.
The Si~{\sc iii}~$\lambda$4560 line in HVG SNe Ia is shallower
than that of LVG SNe (\eg \cite{Pignata2008}). 
\citet{Tanaka2008} synthesized
spectra for several HVG and LVG SNe Ia at earliest phases 
($t\lesssim -1$ week) and suggested that the photospheric
temperature of HVG SNe Ia is lower than those of LVG SNe Ia.
As shown in figures 4 and 5, the absorption line of 
Si~{\sc iii}~$\lambda$4560 in SN~2006X becomes stronger with time.
This fact is consistent with the decreasing trend of $\cal R$(Si~{\sc ii}).
However, the Si~{\sc iii}~$\lambda$4560 is as shallow as that
of other HVG SNe Ia.
This suggests that the photospheric temperature 
is not very high, being similar to other HVG SNe, 
which is in contradiction to the low $\cal R$(Si~{\sc ii}). 

We suggest that the $\cal R$(Si~{\sc ii}) may not be a good 
temperature indicator for the rapidly expanding SNe Ia at the 
premaximum phases.
It is cautioned that  $\cal R$(Si~{\sc ii}) could be contaminated 
by other emission components.
In fact, \citet{Bongard2008} suggested that the Fe~{\sc ii} and 
Fe~{\sc iii} emissions from the inner ejecta form a 
pseudo-continuum around the Si~{\sc ii}~$\lambda$5972 feature,
which could weaken the Si~{\sc ii}~$\lambda$5972 absorption line.
These effects might be strong in the rapidly expanding SNe Ia.
To clarify this issue,
it is necessary to increase the sample of high-velocity SNe Ia
with very early spectroscopic observations.

\section{Conclusions}

 We presented spectroscopic and photometric 
 observations of Type Ia SN 2006X.
 From our premaximum spectra, it is found that the SN is 
 one of the most rapidly expanding Type Ia SNe.
 The extremely high velocities of the Si~{\sc ii}~$\lambda$6355
 and S~{\sc ii}~$\lambda$5640 lines and the weakness of
 O~{\sc i}~$\lambda$7773
 line suggest that an intense nucleosynthesis may take place
 in the outer layers of SN 2006X, 
 consistent with the delayed detonation model.
 Circumstellar interactions might also explain the high 
 velocity features.

 The evolution of the $\cal R$(Si~{\sc ii}) ratio is unique 
 in SN~2006X.
 The $\cal R$(Si~{\sc ii}) is very low throughout the 
 premaximum phases (being unlike the other HVG SNe in
 comparison), which is usually interpreted as a high temperature.
 However, the absorption feature of Si~{\sc iii} 
 $\lambda$4560 is as shallow as those in other HVG SNe,
 suggesting the photospheric temperature is comparable with other HVG SNe.
 This inconsistency might imply that the 
 Si~{\sc ii}~$\lambda$5972 line
 is weakened by contamination of the underlying emission.
 We suggest that the $\cal R$(Si~{\sc ii}) may not be a good
 temperature indicator for rapidly expanding SNe at premaximum 
 phases.

\vspace{10mm}

 We would like to thank the NHAO and GAO staff for enabling us to 
make frequent spectroscopic observations using the NAYUTA telescope
and the 1.5-m telescope.
 The cooperation by graduate students of Osaka Kyoiku University, 
 Y. Ishii, M. Kamada, S, Mizoguchi, S. Nishiyama, N. Sumitomo and 
 K. Tanaka in performing the photometric observations is 
 gratefully acknowledged. 
 We thank an anonymous referee for many comments which helped us in
 improving this paper.
 M.T. is supported by the JSPS 
 (Japan Society for the Promotion of Science) 
 Research Fellowship for Young Scientists. \\\\\\

\clearpage
\begin{figure}
 \begin{center}
  \FigureFile(140mm,140mm){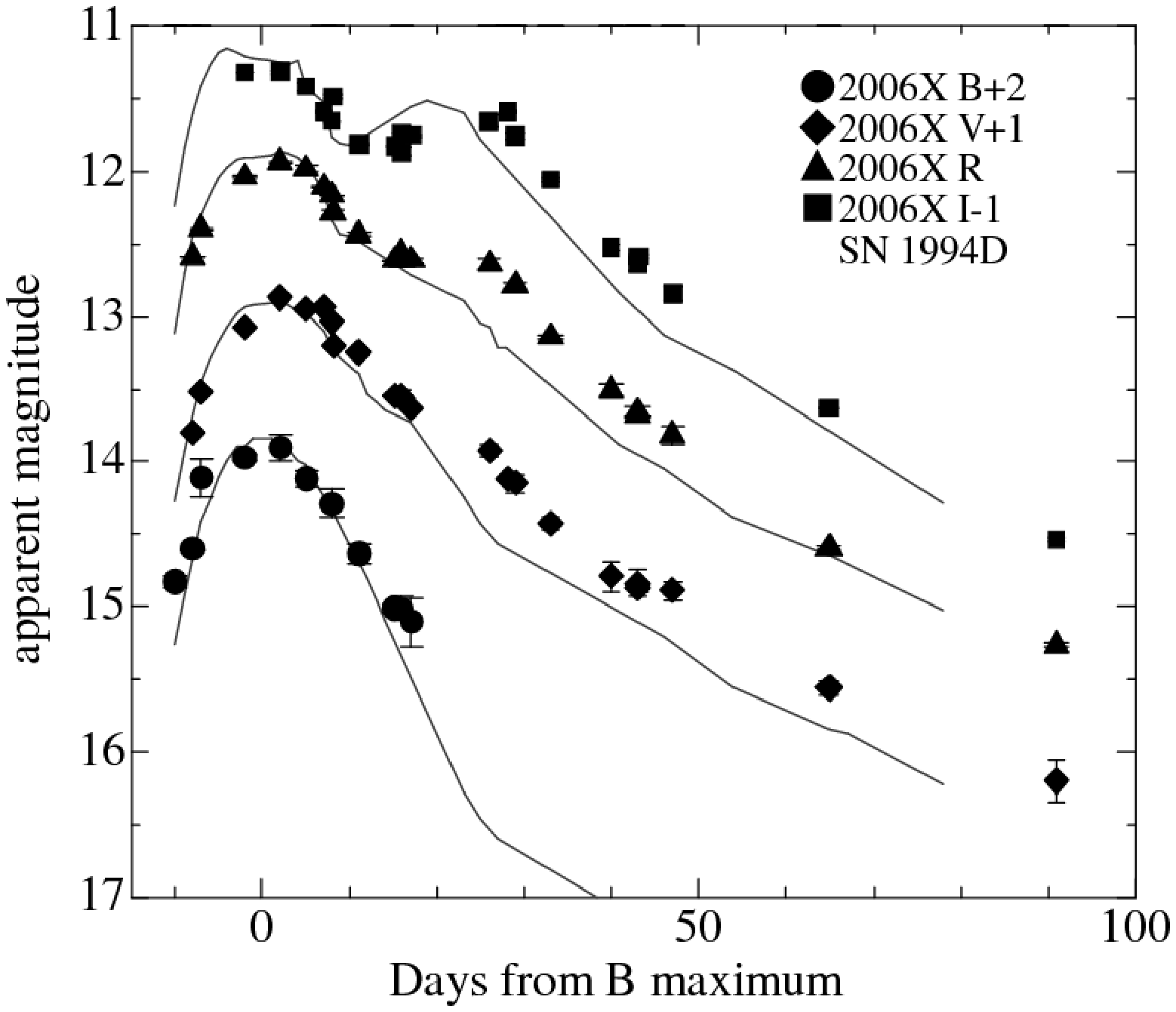}
 \end{center}
 \caption{Light curves of SN~2006X in $B$, $V$, $Rc$ and $Ic$ bands
 (symbols) compared with those of 
 normal SN Ia 1994D (lines, \cite{Patat1996}).
 The extinction in the host and our galaxies has been corrected
 (see \S 3.2). }
    \end{figure}
\clearpage
\begin{figure}
 \begin{center}
  \FigureFile(140mm,140mm){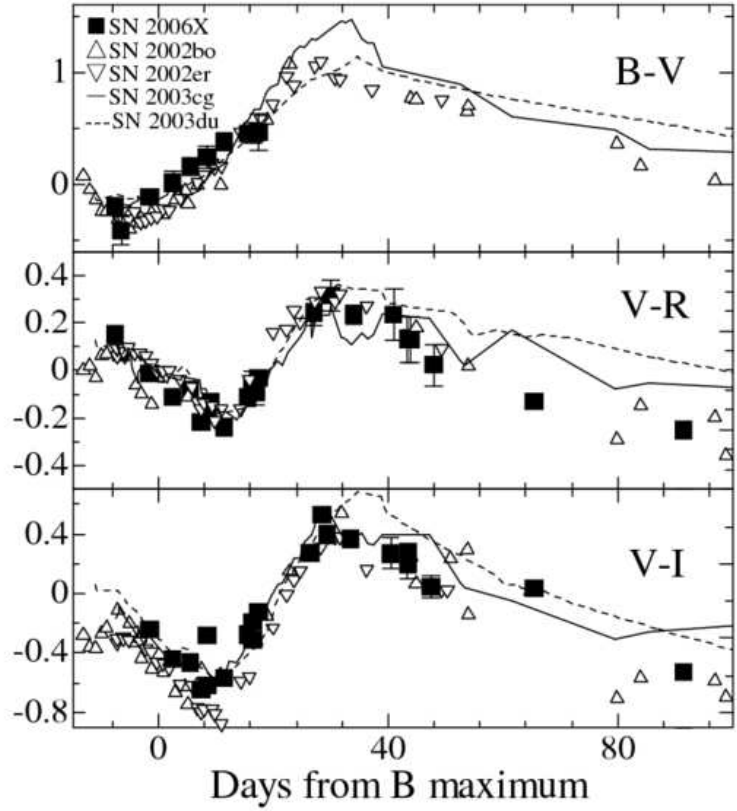}
 \end{center}
 \caption{Reddening-corrected color curves of SN~2006X compared 
   with SNe 2002bo, 2002er, 2003cg and 2003du 
   \citep{Krisciunas2004, Pignata2004,
   Elias2006,Stanishev2007}. 
   The color curves of those comparison SNe are shifted to  
   match the colors of SN 2006X at maximum.}
\end{figure}

\clearpage
\begin{figure}
 \begin{center}
  \FigureFile(140mm,140mm){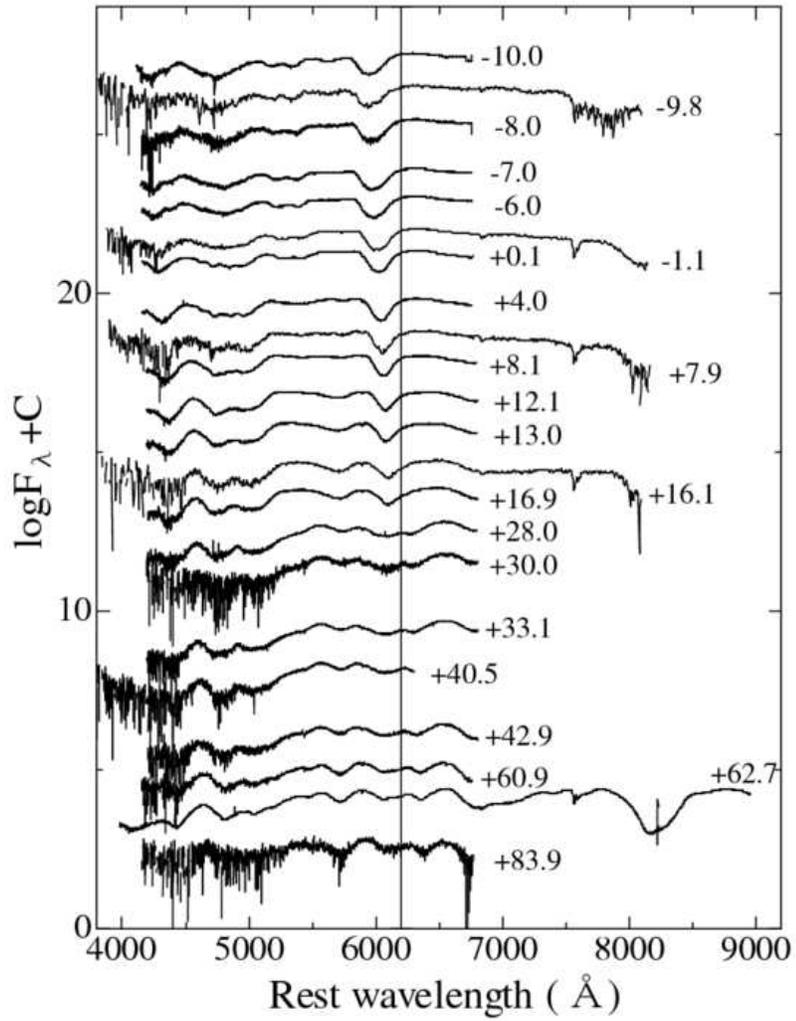}
 \end{center}
 \caption{Spectral evolution of SN 2006X from $t=-10.0$ d 
 to +83.9 d. 
 A vertical line is drawn at 6200 \AA \ to see
 the evolution of Si~{\sc ii} $\lambda$6355 feature.}
\end{figure}

\clearpage
\begin{figure}
 \begin{center}
  \FigureFile(140mm,140mm){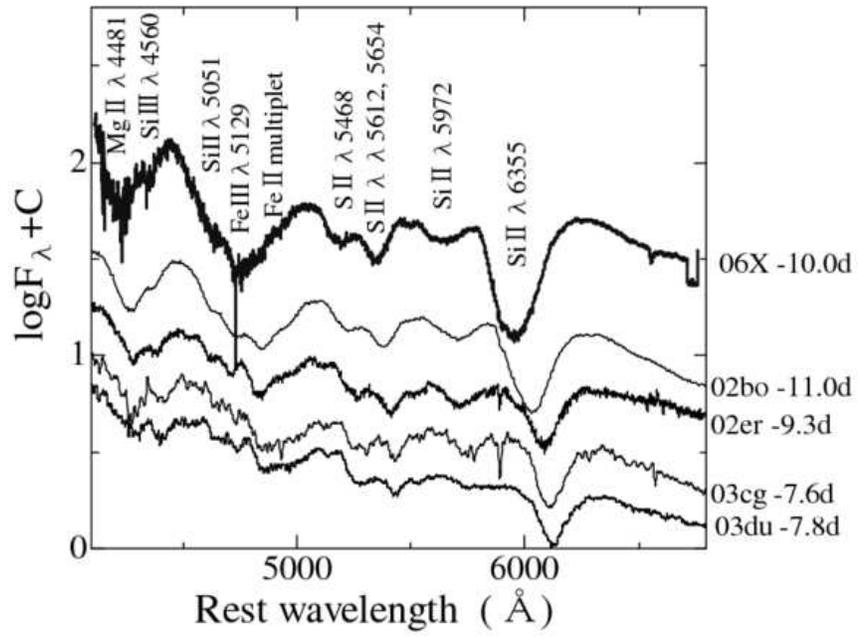}
 \end{center}
 \caption{Spectrum of SN 2006X at $t=-10.0$ d is compared with 
 those of SNe Ia 2002bo, 2002er, 2003cg, and 2003du 
 at similar epochs. The reddening has been corrected.}
\end{figure}
\clearpage
\begin{figure}
 \begin{center}
  \FigureFile(160mm,160mm){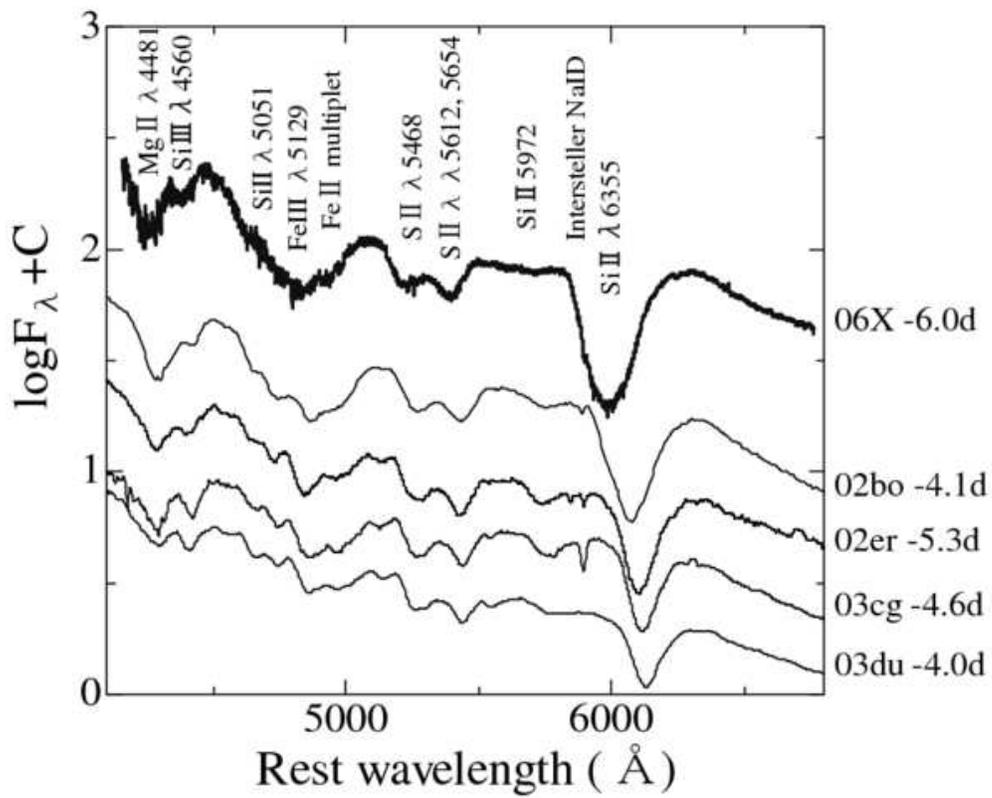}
 \end{center}
 \caption{
   Same as figure 4, but around $t=-6$ d.}
\end{figure}

\clearpage
\begin{figure}
 \begin{center}
  \FigureFile(140mm,140mm){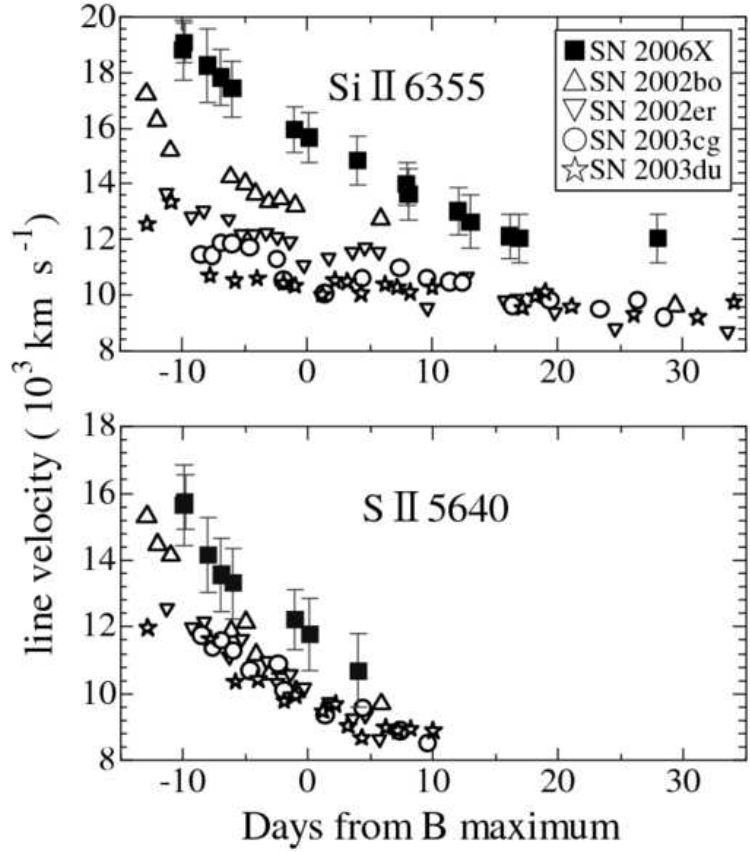}
 \end{center}
 \caption{Time evolution of line velocity of  
 Si~{\sc ii}~$\lambda$6355 (upper panel) and 
 S~{\sc ii}~$\lambda$5640 (lower panel).
 The error bar is estimated by the root sum square of the $1\sigma$ error
 of the velocity measurements by Gaussian fitting and 
 the wavelength resolution of our spectroscopy (see \S 2.2).}
\end{figure}

\clearpage
\begin{figure}
 \begin{center}
  \FigureFile(140mm,140mm){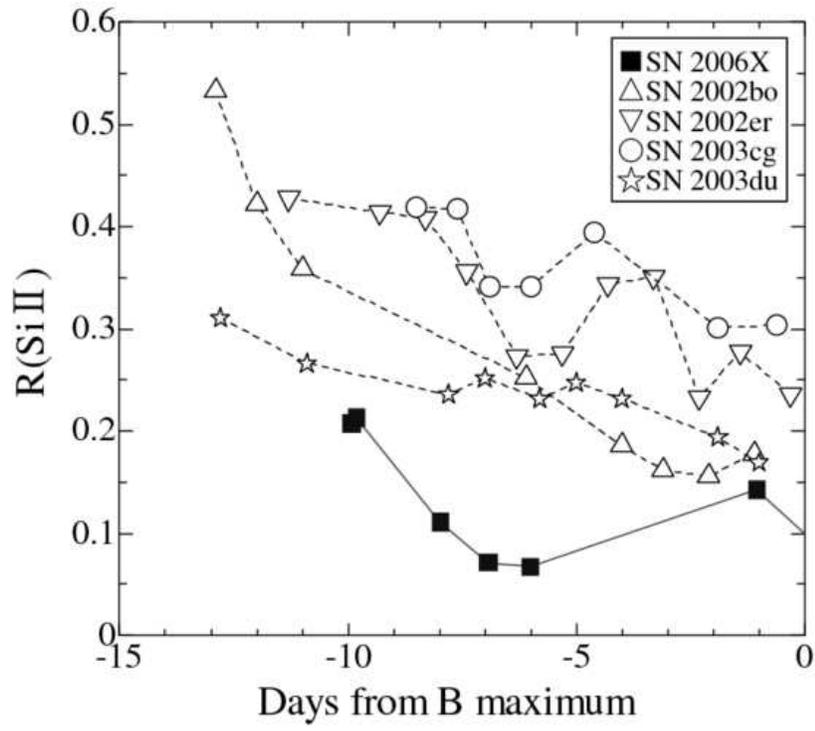}
 \end{center}
 \caption{Time evolution of $\cal R$(Si~{\sc ii}), the depth ratio of
 Si~{\sc ii}~$\lambda$5972 to Si~{\sc ii}~$\lambda$6355, before
 maximum brightness.}
\end{figure}

\clearpage
\begin{table}
\caption{The $BVRcIc$ photometry for SN~2006X}
\begin{center}
\begin{tabular}{cccccccc}
\hline \hline
Date       & MJD      & Epoch$^{\rm a}$ & $B$      & $V$      & $Rc$  & $Ic$ & Obs. \\
\hline
2006/02/09 & 53775.69 & -9.98 & 16.25 $\pm$0.05 &              &              &     & NHAO \\
2006/02/11 & 53777.61 & -8.06 & 16.03 $\pm$0.13 & 14.88 $\pm$0.03 & 14.09 $\pm$0.02 &     & NHAO \\
2006/02/12 & 53778.65 & -7.02 & 15.54 $\pm$0.13 & 14.61 $\pm$0.01 & 13.90 $\pm$0.01 &     & NHAO \\
2006/02/17 & 53783.61 & -2.06 & 15.39 $\pm$0.02 & 14.16 $\pm$0.01 & 13.54 $\pm$0.00 & 13.17 $\pm$0.01 & NHAO \\
2006/02/21 & 53787.65 &  1.98 & 15.32 $\pm$0.09 & 13.96 $\pm$0.02 & 13.44 $\pm$0.00 & 13.16 $\pm$0.01 & NHAO \\
2006/02/24 & 53790.66 &  4.99 & 15.54 $\pm$0.05 & 14.04 $\pm$0.01 & 13.48 $\pm$0.02 & 13.27 $\pm$0.03 & NHAO \\
2006/02/26 & 53792.73 &  7.06 &              & 14.03 $\pm$0.01 & 13.61 $\pm$0.01 & 13.44 $\pm$0.01 & NHAO \\
2006/02/27 & 53793.65 &  7.98 & 15.71 $\pm$0.09 & 14.13 $\pm$0.02 & 13.67 $\pm$0.00 & 13.51 $\pm$0.01 & NHAO \\
2006/02/27 & 53793.77 &  8.10 &              & 14.29 $\pm$0.01 & 13.78 $\pm$0.02 & 13.34 $\pm$0.01 & OKU \\
2006/03/02 & 53796.72 & 11.05 & 16.06 $\pm$0.08 & 14.34 $\pm$0.02 & 13.94 $\pm$0.02 & 13.67 $\pm$0.01 & NHAO \\
2006/03/06 & 53800.80 & 15.13 & 16.43 $\pm$0.02 & 14.64 $\pm$0.01 & 14.11 $\pm$0.00 & 13.68 $\pm$0.01 & NHAO \\
2006/03/07 & 53801.56 & 15.89 & 16.44 $\pm$0.08 & 14.64 $\pm$0.04 & 14.06 $\pm$0.00 & 13.60 $\pm$0.01 & NHAO \\
2006/03/07 & 53801.61 & 15.94 &              & 14.65 $\pm$0.05 & 14.10 $\pm$0.02 & 13.72 $\pm$0.03 & OKU \\
2006/03/08 & 53802.76 & 17.09 & 16.52 $\pm$0.17 & 14.72 $\pm$0.01 & 14.11 $\pm$0.02 & 13.60 $\pm$0.01 & NHAO \\
2006/03/17 & 53811.62 & 25.95 &              & 15.02 $\pm$0.04 & 14.14 $\pm$0.04 & 13.51 $\pm$0.04 & OKU \\
2006/03/19 & 53813.81 & 28.14 &              & 15.21 $\pm$0.02 &        & 13.44 $\pm$0.00 & NHAO \\
2006/03/20 & 53814.62 & 28.95 &              & 15.24 $\pm$0.06 & 14.29 $\pm$0.03 & 13.61 $\pm$0.02 & OKU \\
2006/03/24 & 53818.64 & 32.97 &              & 15.52 $\pm$0.04 & 14.64 $\pm$0.01 & 13.91 $\pm$0.02 & OKU \\
2006/03/31 & 53825.61 & 39.94 &              & 15.88 $\pm$0.10 & 15.01 $\pm$0.04 & 14.37 $\pm$0.02 & OKU \\
2006/04/03 & 53828.58 & 42.91 &              & 15.93 $\pm$0.09 & 15.16 $\pm$0.04 & 14.49 $\pm$0.04 & OKU \\
2006/04/03 & 53828.71 & 43.04 &              & 15.96 $\pm$0.04 & 15.18 $\pm$0.00 & 14.43 $\pm$0.01 & NHAO \\
2006/04/07 & 53832.63 & 46.96 &              & 15.98 $\pm$0.06 & 15.32 $\pm$0.06 & 14.70 $\pm$0.04 & OKU \\
2006/04/25 & 53850.54 & 64.87 &              & 16.65 $\pm$0.05 & 16.10 $\pm$0.01 & 15.48 $\pm$0.01 & NHAO \\
2006/05/21 & 53876.62 & 90.95 &              & 17.29 $\pm$0.15 & 16.76 $\pm$0.02 & 16.39 $\pm$0.01 & NHAO \\

\hline
\end{tabular} \\
$^{\rm a}$ Days measured from the $B$-band maximum, MJD 53785.67
\end{center}
\end{table}

\clearpage
\begin{table}
\caption{Spectroscopic parameters for SN 2006X}
\begin{center}
\begin{tabular}{ccccccccc}
\hline \hline
Date       & MJD      & Epoch$^{\rm a}$ & Exp.$^{\rm b}$  & S~{\sc ii}~$\lambda$5640$^{\rm c}$ 
& Si~{\sc ii}~$\lambda$6355$^{\rm c}$ &
 $\cal R$(Si~{\sc ii}) & Obs.  \\ 
           &          & (days)& (s)  & ($10^{3}$ \kms )
 & ($10^{3}$ \kms )                  &   &   &      \\ 
\hline 
2006/02/09 & 53775.71 & -9.96 & 5400 & $15.6\pm 1.2$ 
 & $18.8\pm 1.1$ & $0.207\pm 0.012$  & NHAO\\
2006/02/09 & 53775.81 & -9.86 &  540  & $15.7\pm 0.8$ 
 & $19.1\pm 0.7$ & $0.213\pm 0.024$ & GAO \\
2006/02/11 & 53777.65 & -8.02 & 1800  & $14.2\pm 1.1$
 & $18.3\pm 1.3$ & $0.111\pm 0.026$ & NHAO\\
2006/02/12 & 53778.71 & -6.96 & 3600  & $13.6\pm 1.1$
 & $17.8\pm 1.0$ & $0.071\pm 0.016$ & NHAO\\
2006/02/13 & 53779.63 & -6.04 & 3600  & $13.3\pm 1.1$
 & $17.4\pm 1.0$ & $0.068\pm 0.011$ & NHAO \\
2006/02/18 & 53784.59 & -1.08 &  900   & $12.2\pm 0.9$
 & $16.0\pm 0.8$ & $0.143\pm 0.024$ & GAO\\
2006/02/19 & 53785.77 &  0.10 & 3600  & $11.8\pm 1.1$ 
 & $15.7\pm 0.9$ & $0.093\pm 0.007$ & NHAO\\
2006/02/23 & 53789.65 &  3.98 & 3600  & $10.7\pm 1.1$
 & $14.9\pm 0.9$ & & NHAO\\
2006/02/27 & 53793.58 &  7.91 & 1500  &               
                & $14.0\pm 0.8$ &  & GAO\\
2006/02/27 & 53793.74 &  8.07 & 1800  &                
                  & $13.6\pm 0.9$ & & NHAO\\
2006/03/03 & 53797.71 & 12.04 & 3600  &                
                & $13.0\pm 0.8$ & & NHAO\\
2006/03/04 & 53798.65 & 12.98 & 1800  &               
                & $12.6\pm 1.0$ & & NHAO\\
2006/03/07 & 53801.80 & 16.13 &  900   &                                & $12.1\pm 0.8$ & & GAO\\
2006/03/08 & 53802.60 & 16.93 & 3600  &                               & $12.0\pm 0.9$ & & NHAO\\
2006/03/19 & 53813.64 & 27.97 & 3600  &                                & $12.0\pm 0.9$ & & NHAO\\
2006/03/21 & 53815.67 & 30.00 & 3600  &  & &  & NHAO\\
2006/03/24 & 53818.73 & 33.06 & 3600  &  & &  & NHAO\\
2006/03/31 & 53826.11 & 40.44 & 3600  &  & &  & NHAO\\
2006/04/03 & 53828.53 & 42.86 & 3600  &  & &  & NHAO\\
2006/04/21 & 53846.55 & 60.88 & 3600 &  & &  & NHAO \\
2006/04/23 & 53848.34 & 62.67 & 1200   & & & & Subaru\\
2006/05/14 & 53869.57 & 83.90 & 7200 &  & &  & NHAO\\
\hline
\end{tabular} \\
 $^{\rm a}$ Days measured from the $B$-band maximum.
 $^{\rm b}$ Total exposure time.\ $^{\rm c}$Line velocity
 measured at the line center (see \S2.2)\
\end{center}
\end{table}

\end{document}